# Practical considerations for specifying a super learner

Rachael V. Phillips[1]*, Mark J. van der Laan[1], Hana Lee[2], Susan Gruber[3]

[1] Division of Biostatistics, School of Public Health, University of California at Berkeley, Berkeley, California, United States of America

[2] Office of Biostatistics, Center for Drug Evaluation and Research, United States Food and Drug Administration, Silver Spring, Maryland, United States of America

[3] Putnam Data Sciences, LLC, Cambridge, Massachusetts, United States of America

* Corresponding author. University of California at Berkeley School of Public Health, 2121 Berkeley Way, Berkeley, California, United States of America, 94704.
Email: rachaelvphillips@berkeley.edu

## Summary

Common tasks encountered in epidemiology, including disease incidence estimation and causal inference, rely on predictive modeling. Constructing a predictive model can be thought of as learning a prediction function, i.e., a function that takes as input covariate data and outputs a predicted value. Many strategies for learning these functions from data are available, from parametric regressions to machine learning algorithms. It can be challenging to choose an approach, as it is impossible to know in advance which one is the most suitable for a particular dataset and prediction task at hand. The super learner (SL) is an algorithm that alleviates concerns over selecting the one "right" strategy while providing the freedom to consider many of them, such as those recommended by collaborators, used in related research, or specified by subject-matter experts. It is an entirely pre-specified and data-adaptive strategy for predictive modeling. To ensure the SL is well-specified for learning the prediction function, the analyst does need to make a few important choices. In this Education Corner article, we provide step-by-step guidelines for making these choices, walking the reader through each of them and providing intuition along the way. In doing so, we aim to empower the analyst to tailor the SL specification to their prediction task, thereby ensuring their SL performs as well as possible. A flowchart provides a concise, easy-to-follow summary of key suggestions and heuristics, based on our accumulated experience, and guided by theory.

> **Key messages**
> - Use the recommended performance metrics provided in Step 1 of the flowchart to select a performance metric that aligns with the intended real-world task and accurately reflects how closely it approximates the true prediction function of interest.
> - Calculate the effective sample size from the analytic dataset by following flowchart Step 2, to use as a simple rule of thumb for defining the cross-validation scheme (Step 3), and to guide the specification of the collection of algorithms included in the SL library (Step 4).
> - Evaluate and construct SL libraries according to the following criteria for a well-specified library: a rich library of algorithms that is diverse in its learning strategies, able to adapt to a range of underlying functional forms for the true prediction function, computationally feasible, and effective at handling high dimensional data.
> - Distinguish between an ensemble SL (eSL) that combines predictions from multiple individual algorithms, and a discrete SL (dSL) that selects a single, best, performer based on cross-validated performance.
> - Justify the use of the dSL as the final model, to select the best performing algorithm, including one or more eSLs alongside each individual algorithm in the original library.

# Introduction

To answer scientific questions and test hypotheses, we curate and learn from data. This often entails using the data to learn a prediction function: a function that takes in covariate data and outputs a predicted value. Prediction is used in epidemiology in a variety of contexts: disease phenotyping, evaluating risk factors for certain health outcomes, mapping the spread of disease, and event forecasting. Prediction is also an integral component in the estimation of causal effects. Traditionally prediction functions were learned by fitting pre-specified parametric regression models to data; however, more flexible learning algorithms have been shown to produce more accurate results. For example, in-hospital mortality prediction scores based on machine learning have been shown to improve upon conventional, logistic regression-based scores [1].

In practice it is difficult to choose a single algorithm (or "learner"). There are many options, and no one is expert in all of them. Moreover, it is impossible to know in advance which learner is most suitable for the particular dataset and prediction task at hand. The super learner (SL) solves the issue of algorithm selection by considering a large set of user-specified algorithms, from parametric regressions to nonparametric machine learning algorithms (e.g., neural nets, support vector machines, and decision and regression trees). It alleviates concerns over selecting the one "right" algorithm while benefiting from considering a diverse set, including those recommended by collaborators, used in related research, or specified by subject-matter experts. The SL is grounded in optimality theory that guarantees for large sample sizes the SL will perform as well as possible, given the specified algorithms considered [2]. The robustness of this entirely pre-specified and data-adaptive approach is supported by numerous practical applications [1, 3–10]. SL plays an important role in targeted maximum likelihood estimation (TMLE), where it is recommended for estimating outcome regressions, propensity scores, and missingness mechanisms.

Existing SL software packages in the R programming language [11], *SuperLearner* [12] and *sl3* [13], make using SL straightforward. By providing a single interface to a rich and diverse variety of algorithms, the analyst is not required to learn the ins and outs of different software packages. Nevertheless, they do need to make a few choices to define a SL that is well-specified for their prediction task. This paper provides an overview of key components of a SL specification, and a flowchart that guides an analyst through the process (Figure 1, on the following page). A glossary of terminology used in the text is also provided (Table 1, on pages 9–10).

## Overview of the super learner

The idea to "lump together [algorithms] into one grand melee" dates back to the early 1960s (George Barnard) [14]. It wasn't until 1992 that a methodology for doing this, stacking, was proposed by David Wolpert [15]. Specifics of the implementation, referred to by Wolpert as an "art" in 1992, became a science in 1996, when Leo Breiman demonstrated the utility of non-negative least squares (NNLS) regression for combining predictions from algorithms fit to the same dataset (meta-learning) [16]. In 2007 previous theory provided by Mark van der Laan, Sandrine Dudoit, and Aad van der Vaart [17–20] was extended, proving that in large samples stacking represented an optimal system for learning [2]. This is when the stacking algorithm took on an additional name, "Super Learner".

The SL learns a function of a specified set, or "library", of algorithms (Figure 2). Each algorithm in the library is considered a "candidate" estimator of the true prediction function. SLs can be grouped into two types, depending on their approach to meta-learning. An ensemble SL (eSL) can use any parametric or nonparametric algorithm to create a function that combines the predictions from each candidate. For example, NNLS produces a weighted sum of predictions from each candidate (Figure 2, on page 11).

Cross-validation (CV) is essential to the SL in estimating how well a trained algorithm performs when making predictions on novel data drawn from the same distribution as the training data. V-fold CV (VFCV) is a splitting of the dataset into V disjoint validation sets and corresponding training sets (the complement of each validation set, Figure 2). For each fold, each candidate is trained on observations in the training set, and then predicted outcomes are obtained for observations in the validation set (Figure 2). A fold-specific evaluation of each trained candidate's predictive performance on the validation set is defined with respect to a risk function, or "performance metric", such as the mean squared error (MSE). The CV performance of a trained candidate learner is its CV risk averaged over all validation sets.

Detailed descriptions outlining the SL procedure are widely available in the literature [21,22]. This paper focuses on practical advice on how to tailor the SL to robustify performance, based on characteristics of the data and the substantive goal. Key components are the library, VFCV scheme, and the dSL's performance metric.

Figure 1. Flowchart for specifying a super learner (SL).

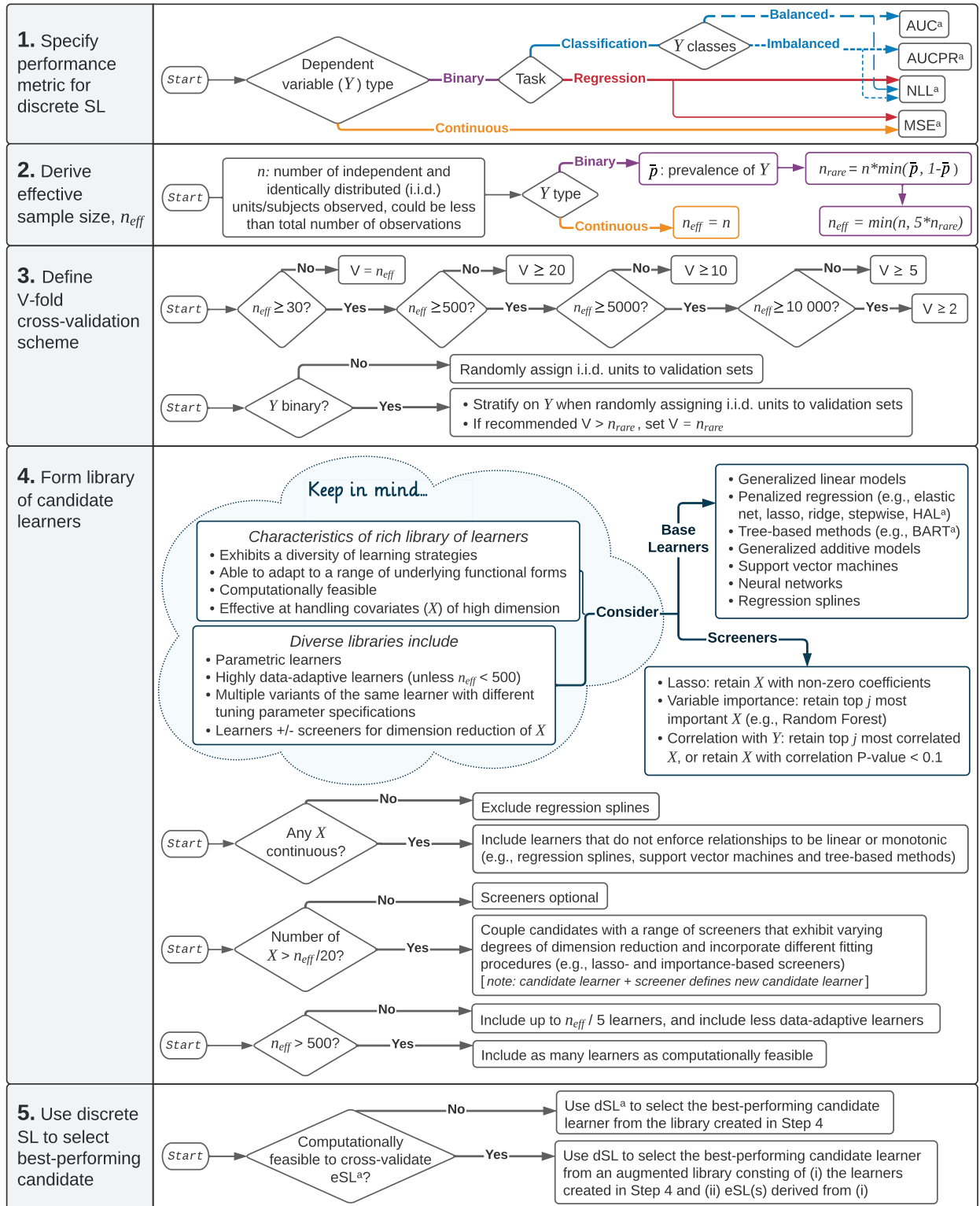

[a] AUC, area under the receiver-operating characteristic curve; AUCPR, area under the precision-recall curve; NLL, negative log-likelihood; MSE, mean squared error; HAL, highly adaptive lasso; BART, Bayesian additive regression trees; eSL, ensemble super learner; dSL, discrete super learner

# Defining a well-specified super learner

In this section, we provide step-by-step guidelines for tailoring the SL specification to perform as well as possible for the prediction task at hand. Our recommendations depend on the information available in the data and the prediction problem. The heuristics discussed here are accompanied by a flowchart (Figure 1). Introductions to the standard SL software (*SuperLearner* and *sl3* R packages), and examples, are freely available online [23,24].

## Preliminaries: Analytic dataset pre-processing

The analytic dataset consists of observations on an outcome ($Y$) and covariates ($X$). SL requires that this dataset does not contain any missing values. When cleaning the data, the analyst should consider omitting observations where the value of $Y$ is an extreme outlier, to avoid unduly influencing the estimation of the prediction function. In high dimensional settings, predictive performance is sometimes improved by reducing the number of covariates. In the pre-processing stage, only outcome-blind dimension reduction strategies may be considered. Examples include removing covariates that subject-matter experts deem irrelevant, removing covariates that are extremely sparse or constant across all observations, and removing one covariate among a pair of highly correlated covariates or creating a single summary measure from a grouping of correlated covariates. Covariate reduction strategies based on the outcome-covariate relationships in the data, such as using lasso [25], should be incorporated within SL's CV procedure (Step 4).

## Flowchart step 1: Specify the performance metric for the discrete super learner

A performance metric, such as MSE, area under the receiver-operating characteristic curve (AUC), or negative log-likelihood (NLL), quantifies the success of an estimated prediction function (i.e., a trained algorithm). The chosen metric should align with the intended real-world use of the predictions, and it should be optimized (minimized or maximized) by the true prediction function. This guarantees that the evaluation corresponds to the trained algorithm's success in approximating the true prediction function. The flowchart lists suitable metrics for different prediction tasks. All of them are available in the standard SL software. Advanced users can define a custom risk function, so long as it is defined by a loss function that is uniformly bounded (easily achieved in practice by truncating predicted values to the range of $Y$) [2].

## Flowchart step 2: Derive the effective sample size

The amount that can be learned from a dataset depends on the complexity of the prediction task and on the amount of information in the data. When the sample size is larger, algorithms can typically learn more than when it is smaller. However, there can be a sparsity of information in the data, even at large sample sizes. For instance, if $Y$ is a rare binary event, then the information content is limited by the size of the minority class. We introduce the effective sample size ($n_{eff}$) as a useful proxy for the information in the data. $n_{eff}$ plays a role in heuristics for tailoring the SL's specifications to robustify performance.

If $Y$ is continuous, we define $n_{eff} = n$, the number of independent and identically distributed (i.i.d.) observations in the dataset. (If observations in the data are clustered, such as repeated measures on the same individual, then $n$ is the number of independent clusters.) If $Y$ is binary, then the amount of information in the data is limited by the number of events or non-events, whichever is the minority ($n_{rare}$). In this case, we take the prevalence into account by defining $n_{eff} = min(n, 5 * n_{rare})$. We will discuss simple rules of thumb for how $n_{eff}$ can impact the number VFCV folds and the choice of base learners included in the library in the following.

## Flowchart step 3: Define the V-fold cross-validation scheme

The CV predictive performance (CV risk) is an estimate of an algorithm's true out-of-sample predictive performance if it were trained on the entire analytic dataset (true risk). VFCV schemes assign every observation to a single validation set and to V-1 training sets, where V is the number of folds. The sample size in each validation set and training set is approximately $n/V$ and $n - (n/V)$, respectively. Figure 2 depicts a VFCV scheme with three folds.

### Selecting the number of folds

V impacts the bias and variance of the estimated CV risk [26]. This estimate more closely approximates the true risk when the distribution of the data in each training set is a faithful representation of the distribution in the full analytic dataset. Ensuring that the joint distribution of covariates in each training set reflects that in the full analytic dataset is accomplished by setting V large. For example, when V is 20, each training set contains 95% of the available observations. V typically ranges between 2 and 20, with larger values recommended for smaller $n_{eff}$.

As V increases from 2 to $n$, the sample size of each training set increases, so the algorithm's fits to each training set approach the full analytic dataset fit. This decreases bias in the CV risk estimate. Also, as V increases from 2 to $n$, the increase in overlap in the training sets increases the correlations among the algorithm's fits to the training data. This tends to increase the variance of the CV risk estimate, which is a mean of these correlated random variables. Theoretical optimality of the SL requires that V increases at a slower rate than $n$; the latter condition rules out leave-one-out CV (LOOCV) for large $n$ [16–18].

We recommend choosing larger V in the SL, keeping computational feasibility in mind. All of this considered, Step 3 of the flowchart recommends LOOCV for very small $n_{eff}$, and large V (smaller than $n_{eff}$) as $n_{eff}$ grows. Reducing V reduces the size of each training set. When $n_{eff}$ is large, this will improve computational feasibility without dramatically biasing the CV risk estimate.

### Stratified cross-validation

When $Y$ is binary, the prevalence of $Y$ in each training set should match the overall prevalence. This can be achieved with stratified VFCV, an option in existing SL software for randomizing the assignment of observations to validation sets within strata of $Y$.

### Cross-validation with clustered data

When the data are not i.i.d., clustered observations must be assigned as a group to the same validation and training sets. This ensures the validation data are completely independent of the training data, and the loss function is evaluated at the cluster level. VFCV with clustered data is specified with existing SL software by supplying a cluster identifier to the "id" argument.

## Flowchart step 4: Form library of candidate learners

An ideal, rich library is diverse in its learning strategies, able to adapt to a range of underlying functional forms for the true prediction function, computationally feasible, and effective at handling high dimensional data. Diverse libraries include parametric learners, highly data-adaptive learners (when $n_{eff}$ is not small), multiple variants of the same learner with different tuning parameter specifications, and learners with a range of screeners for dimension reduction when the data are high dimensional.

### Base learners

Base learners are algorithms that are not fully specified but define a particular learning strategy, such as lasso [25] or random forest [27]. Many base learners are listed at the top right-hand corner of Step 4 in the flowchart. A base learner is a building block for constructing one or more fully specified learners having values provided for all tuning parameters.

Different types of base learners will fit the true prediction function in different ways. When the underlying functional form is simple, each may be able to closely approximate the true prediction function. When it is complex, one type of algorithm may be more successful than another. For example, unlike a main terms parametric model, a tree-based algorithm like random forest [27] inherently models interactions and is unaffected by monotone transforms of the data. Since the true functional form is unknown, it is a good idea to consider a variety of base learners, and to construct multiple variations of the same base learner with different tuning specifications. There is no harm in including a learner that performs poorly in the library, as it will be given a weight of zero by the dSL's meta-learner, and an eSL might use it in combination with predictions from other learners to produce a superior prediction function. This highlights the importance of specifying a rich and diverse library of learners.

### Screener couplings

Covariate screening is essential when the dimensionality of the data is very large, and it can be practically useful in any SL or machine learning application. Screening of $X$ that considers associations with $Y$ must be cross validated to avoid biasing the estimate of an algorithm's predictive performance [28]. Any candidate learner can be paired with a screening algorithm to establish a new candidate in the library. For example, consider pairing a lasso screener with a generalized linear model (GLM) candidate (e.g., logistic regression). In each CV fold, the lasso regression would be fit to the training data. Only the covariates with non-zero coefficients would be passed to the GLM, which would run a regression of $Y$ on the reduced set of covariates in the fold's training data. Covariates retained in each CV fold may vary.

### Smaller effective sample sizes

When $n_{eff}$ is small, algorithms that have more structure and smooth over areas of little support in the data will likely perform better than more data-adaptive learners. Examples include parametric learners, like lasso, and lower-depth tree-based learners. Highly data-adaptive learners, like neural nets with all but the simplest architecture, require more data to fit their parameters well; they are generally less appropriate for smaller $n_{eff}$ due to the limited information in the data.

**Risk functions**

Algorithms that internally optimize risk functions that are not optimized by the true prediction function or are not well aligned with the prediction task may still be included in the library. For example, consider a binary classification prediction problem, where the dSL uses the AUC performance metric. The dSL's winner-take-all meta-learner will select the candidate with the highest CV AUC. The dSL's library of learners can of course contain algorithms whose objective is to optimize the AUC, but it can also contain algorithms that optimize other risk functions, such as the NLL or MSE. In some binary classification problems, NLL-minimizing algorithms can achieve higher AUC than AUC-maximizing algorithms [29]. This further motivates diversifying the library.

**Respecting the statistical model**

The statistical model is a set of probability distributions that could have given rise to the observations in the data. It embodies knowledge regarding the data-generating process (DGP) and is expressed by the analyst through their choice of algorithm(s) to fit to the data. For instance, when a parametric regression is fit to the data, the statistical model is restricted to a small space of parametric probability distributions. On the other hand, a SL with a diverse library of parametric and nonparametric algorithms, each coupled with a range of screeners, lends itself to a much bigger statistical model.

If the DGP does not reside in the statistical model space then the statistical model is misspecified, and this can lead to mild or severe bias in estimates and misleading results. Conveying knowledge about the DGP through the library can mitigate statistical model misspecification with the SL. For example, if it's known that there are interactions among covariates then the analyst can include learners in the library that pick up on that explicitly (e.g., by including in the library a parametric regression learner with interactions specified in a formula) or implicitly (e.g., by including in the library tree-based algorithms that learn interactions empirically). When little is known about the DGP, the library should be as rich and diverse as possible, with respect to $n_{eff}$ and computational feasibility, to accommodate a range of possible underlying functional forms for the true prediction function.

## Flowchart step 5: Use discrete super learner to select the candidate with the best cross-validated predictive performance

When the eSL is used, we recommend it be evaluated as another candidate in a dSL, along with the individual learners from which it was constructed. This allows the eSL's CV performance to be compared to that from the other learners, and for the final SL to be the candidate learner that achieved the lowest CV risk. If the eSL performs better than any other candidate, the dSL will end up selecting the eSL. Another advantage of this approach is that multiple eSLs that use more flexible meta-learner methods (e.g., nonparametric machine learning algorithms like highly adaptive lasso [30]) can be evaluated simultaneously.

# Conclusion

In this work we walked through each of the choices the analyst needs to make to define a SL that is well-specified for the prediction task at hand, including the choice of learners included in the library, VFCV scheme, and the SL's meta-learner and performance metric. We provided data-adaptive recommendations to assist the analyst in making these decisions. The flowchart provides a concise reference for guiding the analyst in how to effectively use SL in practice. For a variety of practical examples considered in the supplementary materials, we use the flowchart to construct well-specified SLs and provide reproducible R code.

# Funding

This project was funded by the United States Food and Drug Administration (US FDA), pursuant to Contract 75F40119C10155. The content is the view of the author(s), and does not necessarily represent the official views of, nor an endorsement by, the US FDA / Health and Human Services, or the US Government.

# Conflicts of interest

None declared.

Table 1. Glossary of terminology used in this paper.

| Term | Definition |
| --- | --- |
| Covariates, predictors ($X$) | The set of variables to use as input for predicting the outcome. |
| Outcome ($Y$) | The dependent variable to predict as a function of $X$. |
| Analytic dataset | The dataset containing observations on $X$ and $Y$. |
| Prediction function | A function that takes in values of input variables and returns a predicted value. When $Y$ is binary, the predicted value is a predicted probability. |
| True prediction function | The prediction function defined with respect to the unknown, underlying probability distribution that generated the data. |
| Statistical model | The set of probability distributions that could have generated the data. It embodies the knowledge regarding the process that gave rise to the data. |
| Algorithm, learner, machine learning algorithm | A set of instructions that define a prediction function estimator when tuning parameters are specified. Estimating the prediction function (i.e., algorithm training/fitting) is an optimization problem; in learning the function of the input variables, the algorithm aims to optimize some performance metric / risk function (e.g., minimize the mean squared error). |
| Base algorithm, base learner | An algorithm that is not fully specified but defines a particular learning strategy (e.g., random forest). A base learner is used as a building block to |

| Term | Definition |
|---|---|
|  | define one or more fully specified learners, i.e., one or more estimators of the true prediction function. |
| Library | The set of specified algorithms that will be considered by the super learner. |
| Candidate, candidate learner | A specified algorithm included in the super learner library, with values provided for all tuning parameters, optionally coupled with a screening algorithm. Candidates are trained to consider $X$ as input variables. |
| Screener, screening algorithm | A function that returns a subset of $X$. A screener can be coupled with a candidate learner to define a new candidate learner that considers the reduced set of screener-returned $X$ as its covariates. |
| Meta-level covariates ($\hat{Y}$) | The set of variables to use as input by the meta-learner for predicting the outcome, which are transformations of $X$ defined by trained candidates applied to $X$. Hence, the "meta" nature of the meta-level covariates: they are covariates of defined by a function of other covariates. Observed values of $\hat{Y}$ are predicted values returned by the trained candidates, given observed values of $X$ (see Figure 2). |
| Meta-learner, meta-learning algorithm | A specified algorithm that is trained to consider $\hat{Y}$ as input variables. Hence, the "meta" nature of the meta-learner: it learns from what is learned by the candidate learners (see Figure 2). |
| Meta-level dataset | The dataset for fitting the meta-learner, containing the cross-validated $\hat{Y}$ values and the corresponding validation set $Y$ values (see Figure 2). |
| Super learner (SL) | Just like any other algorithm, the SL is a prediction function estimator. The fitted SL's input variables are $X$. The SL's estimated prediction function is special in that it has two layers: the inner layer is the set of prediction functions learned by the candidates, and the outer layer is the prediction function learned by the meta-learner (see Figure 2). |
| Discrete SL (dSL) | A SL that uses a winner-take-all meta-learner called the cross-validated selector. The dSL is therefore identical to the candidate with the best cross-validated performance; its predictions will be the same as this candidate's predictions. |
| Ensemble SL (eSL) | A SL that uses any parametric or non-parametric algorithm as its meta-learner. Therefore, the eSL is defined by a combination of multiple candidates; its predictions are defined by a combination of multiple candidates' predictions. (Note that the dSL can be thought of as a highly constrained or superficial type of eSL, in which dSL predictions are a weighted combination of the candidates' predictions, with predictions from the candidate with the best cross-validated performance given weight one and those from all other candidates given weight zero.) |

Figure 2. Illustration of the super learner (SL) with the following specification: V-fold cross-validation with three folds; library with generalized linear model (GLM), highly adaptive lasso (HAL), support vector machine (SVM), and generalized additive model (GAM) candidates; and non-negative least squares (NNLS) regression meta-learner.

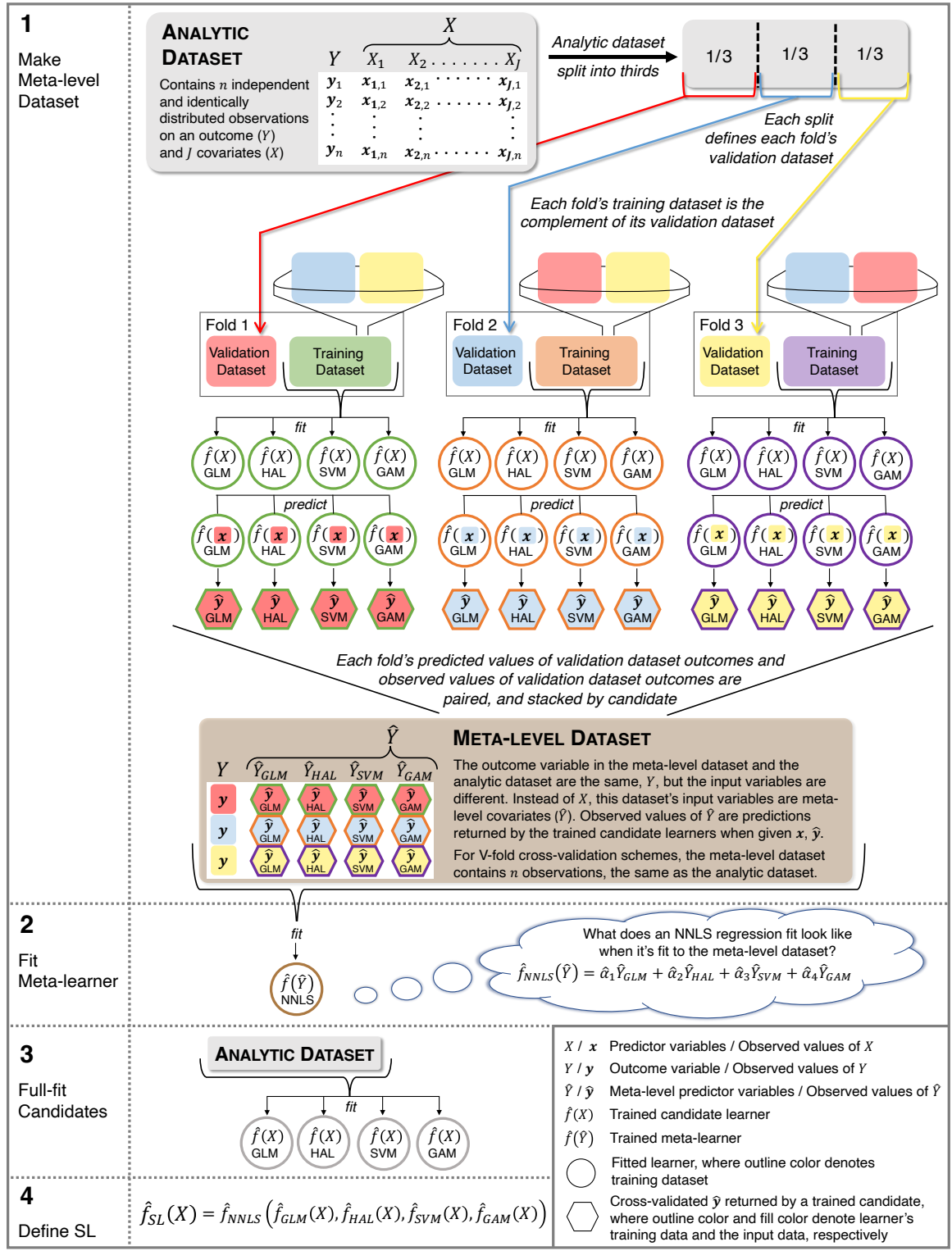